\documentclass[conference]{IEEEtran}
\IEEEoverridecommandlockouts
\usepackage{cite}
\usepackage{amsmath,amssymb,amsfonts}
\usepackage{graphicx}
\usepackage{textcomp}
\usepackage{xcolor}
\ifCLASSINFOpdf
\else
\usepackage[dvips]{graphicx}
\fi
\usepackage{url}
\usepackage{diagbox}
\usepackage[justification=centering]{caption}
\usepackage{graphicx}
\usepackage[noend]{algpseudocode}
\usepackage{algorithmicx,algorithm}
\usepackage{graphicx}
\usepackage{epstopdf}
\usepackage{subfigure}
\usepackage{overpic}
\usepackage{verbatim}
\usepackage{bm}
\usepackage[T1]{fontenc}
\usepackage{siunitx} 
\usepackage{cite}
\usepackage{amsmath,amssymb,amsfonts}
\usepackage{graphicx}
\usepackage{textcomp}
\usepackage{xcolor}
\usepackage{multirow}

\def\BibTeX{{\rm B\kern-.05em{\sc i\kern-.025em b}\kern-.08em
    T\kern-.1667em\lower.7ex\hbox{E}\kern-.125emX}}
\begin{document}

\title{A Privacy-Preserving and End-to-End-Based Encrypted Image Retrieval Scheme\\
}

\author{\IEEEauthorblockN{1\textsuperscript{st} Zhixun Lu}
\IEEEauthorblockA{\textit{College of Information Science} \\
	\textit{and Technology}\\ \textit{Jinan University}
Guangzhou, China \\
luzhixun17@gmail.com}
\and
\IEEEauthorblockN{2\textsuperscript{nd} Qihua Feng}
\IEEEauthorblockA{\textit{School of Computer Science} \\
\textit{and Technology}\\ \textit{Beijing Institute of Technology}
Beijing, China \\
fengqh@bit.edu.cn }
\and
\IEEEauthorblockN{3\textsuperscript{rd} Peiya Li*}
\IEEEauthorblockA{\textit{College of Cyber Security} \\
	\textit{Jinan University}\\
	Guangzhou, China \\
	lpy0303@jnu.edu.cn \\}
}

\maketitle
\footnotetext{*Corresponding author}
\begin{abstract}
	\textbf{Applying encryption technology to image retrieval can ensure the security and privacy of personal images. The related researches in this field have focused on the organic combination of encryption algorithm and artificial feature extraction. Many existing encrypted image retrieval schemes cannot prevent feature leakage and file size increase or cannot achieve satisfied retrieval performance. In this paper, A new end-to-end encrypted image retrieval scheme is presented. First, images are encrypted by using block rotation, new orthogonal transforms and block permutation during the JPEG compression process. Second, we combine the triplet loss and the cross entropy loss to train a network model, which contains gMLP modules, by end-to-end learning for extracting cipher-images' features. Compared with manual features extraction such as extracting color histogram, the end-to-end mechanism can economize on manpower. Experimental results show that our scheme has good retrieval performance, while can ensure compression friendly and no feature leakage.}
\end{abstract}


\IEEEpeerreviewmaketitle

\section{Introduction}

\IEEEPARstart{I}{n} line with the need for growing picture storage and search in a safe and efficient way, encrypted image retrieval has received increasing attention. So far, many cipher-image retrieval schemes (CRSs) have been proposed in the literature. We roughly classify them into two types: encrypting plain-image's features (EPF) and extracting cipher-image's features (ECF). In EPF's schemes \cite{2017EPCBIR,2017AP,2019Secure}, they first extracted features from plain-image, and then applied searchable encryption techniques to protect them, while plain-images were encrypted by traditional cryptographic algorithms. These schemes have a higher level of protection for images and features. However, this two-stage approach (feature extraction \& encryption) which should completed before uploading would incur extra computational cost and inconvenience for users. Thus, some ECF's schemes\cite{2014Histogram,2016AC,2016Markov,Zhang2016Encrypted,2019Huffman,2019Encrypted,2019A,0BOEW,feng2021end} in which features are directly extracted from the cipher-image by the cloud server have been developed. Within the framework of ECF schemes, the user only needs to encrypt the images before uploading, while feature extraction and retrieval are implemented by cloud server. However, these schemes have unresolved disadvantages: first, in order to retrieve images' information effectively, some of these schemes\cite{2014Histogram,2016AC,2016Markov,Zhang2016Encrypted,feng2021end} had the issue of feature leakage; second, some ECF schemes\cite{2014Histogram,2019Huffman} had a significant impact on image compression performance; third, although some existing ECF schemes \cite{2014Histogram,2016AC,Zhang2016Encrypted,2019Huffman,2019Encrypted,2019A,0BOEW} realized cipher-image search effectively, but the performance of retrieval is not ideal.

In this paper, a new end-to-end encrypted image retrieval scheme which follows the framework of ECF is presented. Specifically, we first protect plain-images through block rotation/shuffling and DCT matrix replacement. Then, we employ gMLP network model to learn the high-dimensional features and retrieve the cipher-images by end-to-end. The experiments show that our scheme has good retrieval performance, can ensure compression friendly and no feature leakage.

\section{The Proposed Scheme}

The specific framework of our scheme mainly includes three subjects: image owner, cloud server, and authorized user. First, the image owner uploads and stores the encrypted images in the cloud. Second, authorized user can send query request with encrypted image to server for searching. Third, cloud server measures the similarity between cipher-images in the database and the encrypted query image through trained network model, and then returns the search results to authorized user. Finally, the authorized user obtains the plain-images by using the corresponding key to decrypt cipher-images. It should be noted that if the authorized user is not cipher-image's database uploader, decryption key will be transmitted to the authorized user from image owner through the secure key transmission channel.

This part focuses on the encryption operations used by the image owner, as well as the structure and training of the network model in the server.

\subsection{Implementation of encryption operations}
\begin{figure}[h]
	\centerline{\includegraphics[width=\columnwidth]{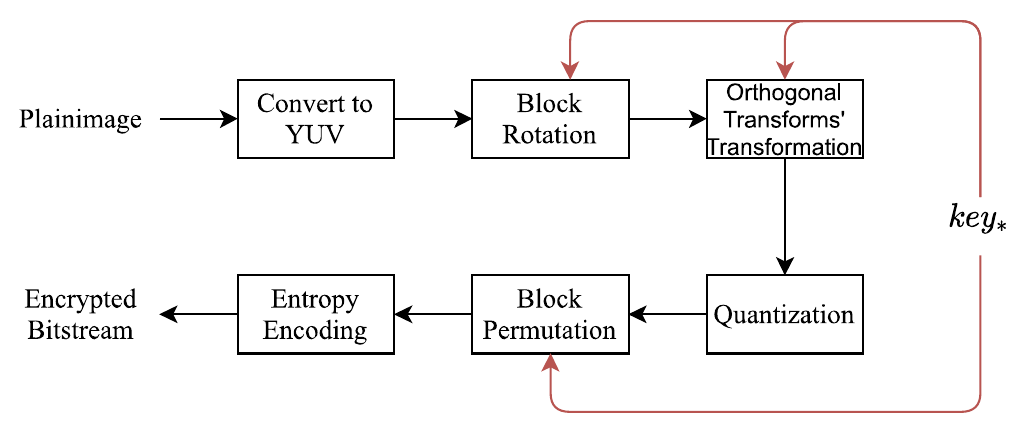}}
	\caption{Diagram of the proposed encryption method.}
	\label{fig.enc_pro}
\end{figure}

Since JPEG images are widely used in our life, it is necessary to study encrypted JPEG image retrieval. Our encryption is realized during JPEG compression process. As presented above, there are three operations in our encryption scheme: block rotation \cite{2019Encryption}, other orthogonal transforms' transformation \cite{2017A} and block permutation. The encryption keys ($\left\{key_*\right\}_{*\in \left\{Y,U,V\right\}}$) are different for each color component and are generated by using SHA-256. Fig. \ref{fig.enc_pro} shows the flowchart of our encryption method.
Image color space is converted from RGB to YUV. Block rotation and orthogonal transforms' transformation are performed for each 8$\times$8 block. Among them, block rotation contains four angles ($\ang{0}, \ang{90}, \ang{180}, \ang{270}$), so it can be controlled by 2 bits selected from $key_*$. After block rotation, we use 63 bits to select 9 transforms from the transform set which includes 128 orthogonal matrices for row's and column's transformation. After that, we quantize all 8$\times$8 blocks with the quantization table of JPEG and apply block permutation to enhance the image security. Finally, encrypted JPEG bitstream can be obtained after zig-zag scan and entroy encoding.




\subsection{Image Retrieval}
\begin{figure*}[h]
	\centering
	\includegraphics[width=0.8\textwidth]{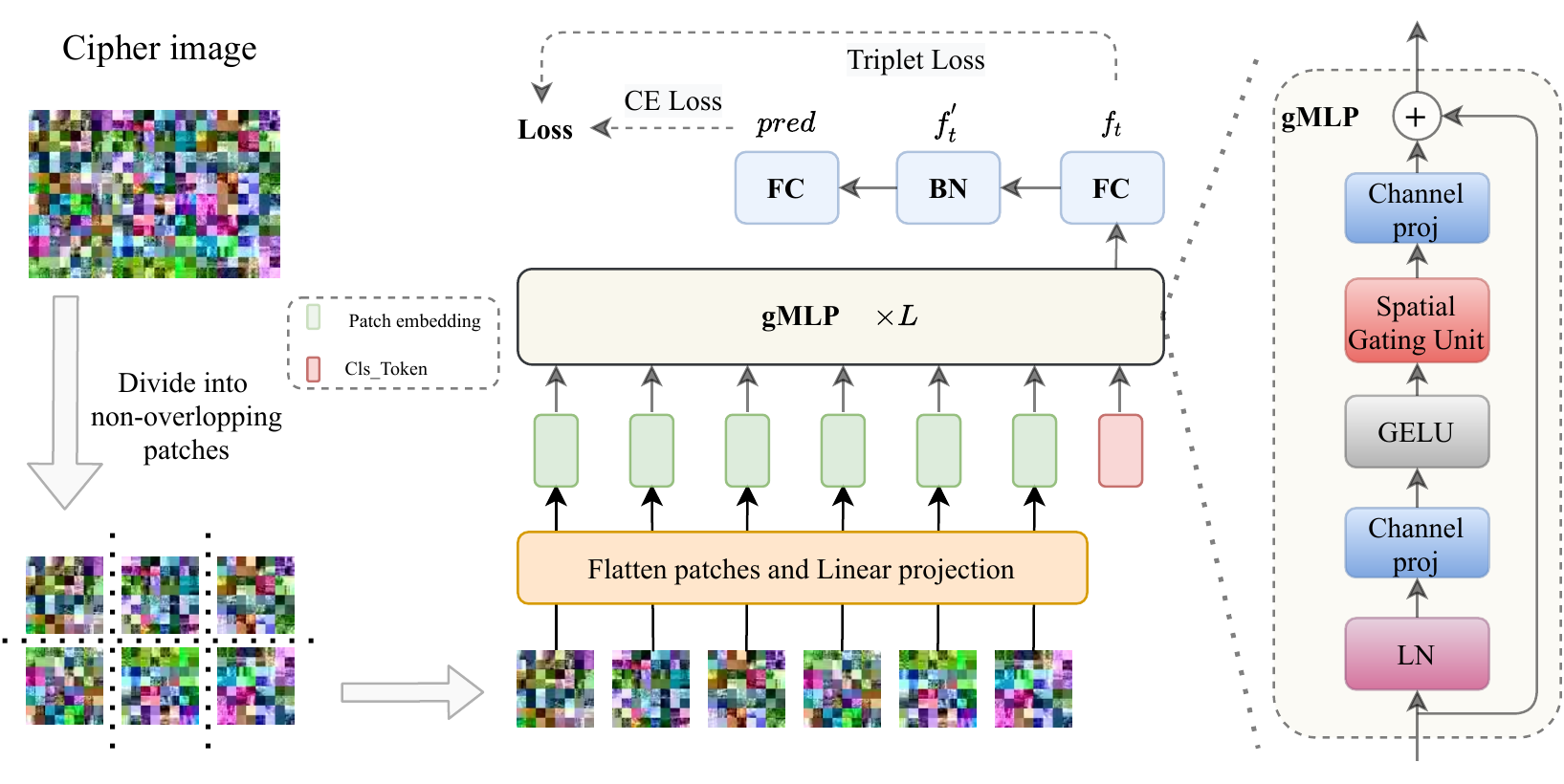}
	\DeclareGraphicsExtensions.
	\caption{Diagram of our network architecture.}
	\label{fig.gmlp}
\end{figure*}

In this paper, we propose to directly extract features from cipher-image by using network model. Cipher-images can be obtained by decoding encrypted JPEG bitstream because of the format-compliant property of our encryption scheme. As shown in Fig.~\ref{fig.gmlp}, we divide the cipher-image into $S$ non-overlapping patches ($P_1, P_2,...,P_{s-1},P_s$). If the image size is $R\times C$, and each patch size is $I\times I$, then the number of patches can be calculate by $S=(R\times C)/I^2$. After that, we flatten patches and map them to $D$ dimensions with a linear projection ($LP$). In addition, we add a learnable embedding $Cls\_Token$ which is developed in \cite{2018BERT} before the first patch as the representation of encrypted image. gMLP developed in \cite{liu2021pay} is a module to enable cross-patch interaction and we apply it to our network model. Suppose $\boldsymbol X$ is the input of gMLP, $\boldsymbol O$ is the ouput of gMLP, the gMLP module can be written as follows:
\begin{equation}
	\boldsymbol U = LP(LN(\boldsymbol X))
\end{equation}
\begin{equation}
	\boldsymbol V = \sigma(\boldsymbol U)
\end{equation}
\begin{equation}
	\boldsymbol O = LP(SGU(V)) + \boldsymbol X
\end{equation}
where $LN$ is Layer Normalization, $\sigma$ is a activation function gelu \cite{hendrycks2020gaussian}. $SGU$ is the spatial gating unit.
After $L$ gMLPs' processing, we can obtain the new cipher-image's representation which is denoted as $Cls\_Token_N$. The $Cls\_Token_N$ is used to implement classification which can be described as follows:
\begin{equation}
	f_t = LP(Cls\_Token_N)
\end{equation}
\begin{equation}
	pred = LP(BN(f_t))
\end{equation}
where $BN$ is the Batch Normalization, $f_t$ is the feature of encrypted image which is used for retrieval step, and $pred$ is the output of our model.

In the training phase, we combine the triplet loss (TL) \cite{schroff2015facenet} and cross entropy loss (CE) to update weights. Triplet loss is commonly used in retrieval schemes, because it can learn the similarity and difference between samples in a good way. The triplet loss can be written as follows:
\begin{equation}
	L_{TL} = max(margin+sim(f_ts, f_tn)-sim(f_ts, f_tp), 0)
\end{equation}
where $sim$ is the Euclidean distance function. We compute the cosine distance to measure the similarity in retrieval stage. $f_ts$ is anchor samples' $f_t$, and $f_tp$/$f_tn$ is the positive/negative samples' $f_t$. The $margin$ is the threshold which is set to 0.3 in our scheme. Cross entropy loss is a commonly used classification loss, and it can be described as follows:
\begin{equation}
	L_{CE} = -\sum_{i=1}^{K} p_{i}log q_{i}
\end{equation}
\begin{equation}
	p_{i} = \begin{cases}
		1, &i=y\cr 0, &otherwise\end{cases}
\end{equation}
where $K$ is the number of classes, $q_{i}$ ($1\leq i\leq K$) is the prediction logit of class $i$ and it can be computed with $fp$, $y$ ($1\leq y\leq K$) is the corresponding label of the sample. We combine above two losses by an add operation, so the final loss of our method can be defined as follows:
\begin{equation}
	L_{A} = L_{TL} + \beta L_{CE}
\end{equation}
where $\beta$ is the parameter which we set to be 1. After we obtain the total loss ($L_{A}$), the Adam optimizer is applied to update the model's weights. 

\section{Experiments}
In this paper, we evaluate our proposed scheme from three aspects: security analysis, compression performance, and retrieval accuracy, and compare the performance with other related works. Corel-1K, Corel-10K\cite{2000SIMPLIcity} and UCID\cite{2003UCID} image databases are used for test. Among them, Corel-10K/Corel-1K contians 10000/1000 JPEG images, and 100 images for each category. UCID database has 1338 uncompressed color images. We test the retrieval accuracy with Corel-10K, and the compression performance with UCID. Corel-1K was used for statistical model-based attack analysis.   

\subsection{Security analysis}\label{AA}
\begin{figure}[h]
	\centering
	\subfigure[]{\includegraphics[width=0.4\columnwidth]{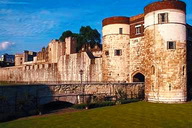}%
		\label{plainimage}}
	\hfil
	\subfigure[]{\includegraphics[width=0.4\columnwidth]{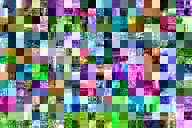}%
		\label{cipherimage}}
	\hfil
	\caption{Encryption example: (a) Plain-image; (b) Cipher-image.}
	\label{image}
\end{figure}

In this section, two types of attacking methods are analyzed: brute-force attack and statistical model-based attack. In Fig.~\ref{image}, the `architecture' image is encrypted using our proposed algorithm as an encryption example. It can be seen that obtaining plain-image's information from the cipher-image is very difficult.
\begin{itemize} 
	\item \textit{Brute-force attack:} This attack method depends on the key space rather than a particular encryption algorithm. Therefore, it should be large enough for the key space of the cryptosystem. In our scheme, there are three main encryption operations: block rotation($N_r$), orthogonal transforms' transformation($N_o$) and block permutation($N_p$). The final key space of our scheme is given by
	
	\begin{equation}
		\begin{aligned}
		N_A(n) &= N_r(n)\times N_o(n)\times N_p(n) \\
		& = 4^n \times (\mathrm{C}_{128}^9)^{n} \times n!
		\end{aligned}
	\end{equation}
	
	
	
	
	
	
	
	
	
	where $n$ is the number of 8$\times$8 block. If image size is 192$\times$128$\times$3, $n$ is 384 for Y component, and 96 for U/V component. The final key space of our encryption scheme is $N_A(n)=N_r(n)\times N_o(n)\times N_p(n)=4^{384+96+96}\times (\mathrm{C}_{128}^9)^{384+96+96} \times 384! \times 96! \times 96!$ which is large enough to resist the brute-force attack.
	
	\item \textit{Statistical model-based attack:} To resist this attack, the plain-image's histogram and cipher-image's histogram should be different.
	Some encryption schemes for JPEG images result in a noticeable information leakage of plain-image \cite{2014Histogram,2016AC,Zhang2016Encrypted,2016Markov}. In Table \ref{Edist}, we calculate the Euclidean distances of four types histograms between original JPEG image and encrypted JPEG image, which are DC coefficient histogram (DCC), DC huffman code histogram (DCH), AC coefficient histogram (ACC) and AC huffman code histogram (ACH). We randomly select 10 images from each category of Corel-1K, thus a total 100 images are used for testing. In theory, distance 0 means that the histogram information is leaked. In Table \ref{Edist}, all distances of our method are not 0 which indicates no information leakage. On the contrary, other schemes\cite{2014Histogram,2016AC,Zhang2016Encrypted,2016Markov} had different degrees of feature leakage. In addition, since the scheme in \cite{2016AC} had the problem of JPEG format incompatible, their distances of DCC and DCH are $\times$s in Table \ref{Edist}. 
	\begin{table}[ht]
		\caption{The euclidean distances between histograms of different methods}
		\renewcommand{\arraystretch}{2}
		\setlength\tabcolsep{5pt}
		\centering
		\begin{tabular}{c|c|c|c|c}
			\hline
			Methods   & DCC      & DCH      & ACC     & ACH     \\ \hline
			Zhang\cite{2014Histogram}     & 0        & 0.30072  & 0       & 0.09920 \\ \hline
			Cheng\cite{2016AC}     & $\times$ & $\times$ & 0       & 0       \\ \hline
			Cheng\cite{Zhang2016Encrypted} & 0.08866  & 0        & 0       & 0       \\ \hline
			Cheng\cite{2016Markov}& 0.01165 & 0 & 0.09069 & 0 \\ \hline
			Ours      & 0.05088  & 0.36284  & 0.00896 & 0.02419 \\ \hline
		\end{tabular}
		\label{Edist}
	\end{table}

\end{itemize}

\subsection{Compression Performance}
\begin{figure}[h]
	\centering
	\includegraphics[scale=0.45]{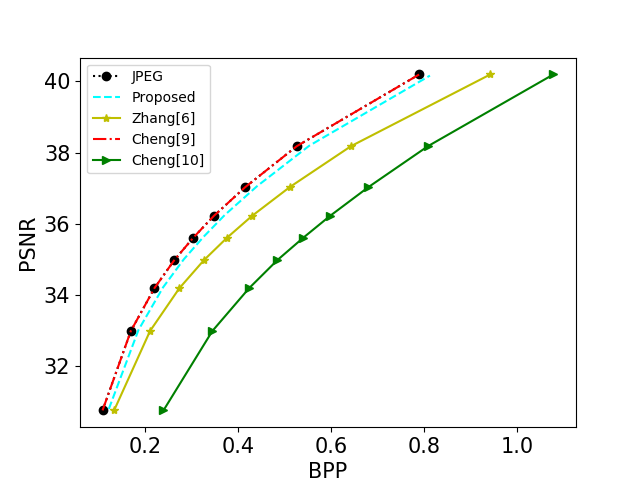}
	\DeclareGraphicsExtensions.
	\caption{Comparison of BPP-PSNR curves for different encryption schemes.}
	\label{fig.psnr}
\end{figure}
In encrypted JPEG image retrieval scheme, if the encryption algorithm results in dramatic damage to compression efficiency, it will cause huge transmission bandwidth and storage space. We calculate the peak signal-to-noise ratio (PSNR) value and bit per pixel (BPP) value of decrypted images to appraise the compression performance of our method. 100 uncompressed images are randomly selected from UCID database as test samples to display the average BPP-PSNR curves of different schemes when QF is 10 to 90. From the BPP-PSNR curves plotted in Fig.~\ref{fig.psnr}, it is clear that when the key is available to the decoder, our scheme' curve is much closer to JPEG compared with that of \cite{2014Histogram} and \cite{2019Huffman}, which indicates better compression performance. Since Cheng et al. \cite{2016Markov}'s scheme protected plain-image by using stream encryption for DCT VLE (Variable Length Encoding), the compression performance of their cipher-image
did not change. However, their scheme had the issue of huffman code histogram leakage.

\subsection{Retrieval Accuracy}
We divide Corel-10K into training set and testing set through random stratified sampling, and the training set contains 7000 images while the testing set contains 3000 images. All methods are evaulated by using the same testing set. For training, we set the batch size to be 50, set epochs as 70 as well as set patch size as 8. To learn better for the model, we use a warmup strategy to bootstrap the network \cite{luo2019bag} in training stage. 

We utilize the mean average precision (mAP) \cite{2007Object} to illustrate the retrieval performance. As shown in Table \ref{mAP}, compare with unsupervised methods \cite{2014Histogram,2019Huffman,0BOEW}, our scheme's mAP is 0.3172 which has been improved more than 0.16, and it means that applying our trained network model to retrieval has a good performance. In schemes related to supervised learning, our scheme's mAP is only about 0.05 higher than Cheng et al. \cite{2016Markov}, but the use of an end-to-end retrieval mechanism in our method can save more manpower than \cite{2016Markov}. In addition, the end-to-end scheme proposed by Feng et al.\cite{feng2021end} has good retrieval performance, but it needs to retain certain blocks from the plain-image, which may result in the leakage of plain-image information.

\begin{table}[h]
	\renewcommand{\arraystretch}{2}
	\setlength\tabcolsep{3pt}
	\centering
	\caption{Retrieval accuracy of different methods}
	\label{mAP}
	\begin{tabular}{c|cccccc}
		\hline
		\multirow{3}{*}{Schemes} & \multicolumn{4}{c|}{No End-to-End}                                                                                                            & \multicolumn{2}{c}{End-To-End}             \\ \cline{2-7} 
		& \multicolumn{3}{c|}{Unsupervised}                                                                         & \multicolumn{3}{c}{Supervised}                                                 \\ \cline{2-7} 
		& \multicolumn{1}{c|}{Zhang{[}6{]}} & \multicolumn{1}{c|}{Cheng{[}10{]}} & \multicolumn{1}{c|}{Xia{[}13{]}} & \multicolumn{1}{c|}{Cheng{[}9{]}} & \multicolumn{1}{c|}{Feng{[}14{]}} & Ours   \\ \hline
		mAP                      & \multicolumn{1}{c|}{0.1478}       & \multicolumn{1}{c|}{0.1008}        & \multicolumn{1}{c|}{0.1267}      & \multicolumn{1}{c|}{0.2651}       & \multicolumn{1}{c|}{0.4342}       & 0.3172 \\ \hline
	\end{tabular}
\end{table}
The number of gMLPs ($L$) and the $f_t$'s dimension ($DIM$) are two parameters which impact the retrieval accuracy of the proposed scheme. We test our scheme by using the same training set and testing set under a range of parameter settings. As shown in Table \ref{dimandn}, the mAP achieves the highest value when $L$ set to be 12 and $DIM$ set to be 128. In addition, we test the retrieval performance of different patch sizes in the network model. According to the Table \ref{patchsize}, the retrieval performance is the best when the patch size is set to be 8$\times$8.

\begin{table}[h]
	\caption{mAP of different $L$ and $DIM$}
	\renewcommand{\arraystretch}{2}
	\setlength\tabcolsep{3pt}
	\centering
	\begin{tabular}{c|c|c|c|c}
		\hline
		\diagbox[width=18mm, height=11mm]{$DIM$}{$L$} & 6 & 8 & 10 & 12   \\ 
		\hline
		128    & 0.2484 & 0.2741 & 0.2987 & \textbf{0.3172}  \\ \hline
		192    & 0.2550 & 0.2818 & 0.3031 & 0.2958   \\ \hline
		256    & 0.2521 & 0.2888 & 0.3031 & 0.2958   \\\hline
		384    & 0.2635 & 0.2732 & 0.2866 & 0.2975   \\
		\hline
	\end{tabular}
	\label{dimandn}
\end{table}

\begin{table}[h]
	\renewcommand{\arraystretch}{2}
	\setlength\tabcolsep{3pt}
	\centering
	\caption{Retrieval accuracy of different patch sizes}
	\begin{tabular}{c|c|c|c}
		\hline
		\diagbox[width=30mm, height=10mm]{Performance}{Patch size} & 8$\times$8 & 16$\times$16 & 32$\times$32 \\ \hline
		mAP  & \textbf{0.3172} & 0.2824  & 0.3046\\ \hline
		Recall@10  & \textbf{0.1526} & 0.1416 & 0.1478\\ \hline
		Recall@30  & \textbf{0.3370} & 0.3097 & 0.3301\\
		\hline
	\end{tabular}
	\label{patchsize}
\end{table}

\section{Conclusion}
In this paper, we propose a new encrypted JPEG image retrieval scheme that extracts features through network model rather than manual work. For encryption scheme, other orthogonal transforms are used to replace 8$\times$8 DCT. To further enhance security, block rotation and block permutation are applied before and after transformation stage, respectively. Experiments show that our encryption algorithm can ensure no information leakage, and maintain good compression performance. For retrieval part, we use a network model containing gMLP which is a strong module to enable cross-patch interaction. Moreover, we add a learnable embedding $Cls\_Token$ for cipher-image representation before the first flattened patch. To better update weights, we combine the triplet loss and cross entropy loss for training. Because we use end-to-end mechanism to extract features, our scheme can save the manpower. The final retrieval performance is good through using our trained model. 

In our next work, we plan to develop other encryption algorithms and consider other image formats. In addition, new techniques will be exploited to extract cipher-image's feature effectively, not just neural network. 


\vspace{12pt}
\color{red}

\end{document}